# An Efficient Scheme for Sensitive Message Transmission using Blind Signcryption

Amit K Awasthi
Department of Applied Science
Hindustan College of Science & Technology,
Farah, Mathura – 281122 (UP) - INDIA,
E-mail: awasthi_hcst@yahoo.com

Sunder Lal
Department of Mathematics,
IBS, Khandari, Agra – 282002(UP) – INDIA

## 1. Introduction

Y. Zheng introduced a new cryptographic primitive Signcryption in [12], which combines digital signature function with a symmetric key encryption algorithm. A digital signature is used for authentication of message and an encryption scheme is used for the confidentiality of messages. A signcryption offers these two properties at the same time and more efficiently. Computational cost in Signcryption is considerably reduced as compared to signature-then-encryption process.

The concept of blind signature was introduced by D Chaum [2]. Blind signature schemes [4–6, 9] are protocols that guarantee anonymity of the participants. A blind signature scheme allows a user to obtain a valid signature for a message from a signer, without signer's seeing the message or its signature. Though it can be verified that the signature is genuine but signer can not link the message signature pair to that instance of signing protocol which has led to this pair. It realizes secure electronic payment systems and protect customer's privacy as well as participants' anonymity.

In this paper, we present a Blind Signcryption Scheme that combines the functionality of blind signature and encryption. The Blind Signcryption ensures anonymity, untracebility and unlinkability.

## 2. Preliminaries

A digital signcryption scheme [12, 13] is cryptographic method that fulfills both, the function of secure encryption and digital signature but at a cost lower than that of signature-then-encryption. In cryptographic notation signcryption consists of a pair of polynomial time algorithms (S, U), where S is the signcryption algorithm, while U is unsigncryption algorithm. S is in general probabilistic, but U is most likely to be deterministic. (S, U) satisfies the following conditions [3].
**Unique unsigncryptability** - The signcryption algorithm S takes the given message m as input and outputs a ciphertext c. On input c the unsigncryption algorithm U recovers the original message uniquely.
**Security** - The protocol SC = (S, U) satisfies simultaneously, the property of digital signature and encryption both. In other words ensures the following:

> *Unforgeability* - It is computationally infeasible for an adaptive attacker to masquerade the signer and create a signcrypted text.

*Nonrepudiation* - Signer cannot deny that he/she is the originator of the signcrypted text with the recipient. In case of dispute it is feasible for some third party to decide it.

*Confidentiality* - No third party can obtain any information about message contents from the content of a signcrypted text.

**Efficiency** - The computational cost of SC and communication overhead of the scheme is smaller than required by signature-than-encryption process.

Before introducing our Blind Signcryption Scheme, we discuss in the following subsections shortened digital signature, Zheng's signcryption scheme.

## 2.1 Shortened Digital Signature standard

As the name suggests it is a shortened form of DSS [7]. Zheng used it for construction of an efficient signcryption scheme.

The parameters used in scheme are:
- $p$: a large prime
- $q$: a large prime factor of $(p-1)$
- $g$: an element of $Z_p^*$ of order $q$
- $h(.)$: a one way secure hash function
- $x_A$: signer A's private key
- $y_A$: signer A's public key, where $y_A = g^{x_A} \mod p$

**Signature Generation**

To sign a message $m$ the Signer A randomly chooses a number $k \in Z_p^*$ and computes

$$r = h(g^k \mod p, m) \qquad (1)$$

and
$$s = \frac{k}{r + x_A} \mod q. \qquad (2)$$

A sends $(m, r, s)$ to B as signature tuple.

**Signature Verification**

For verification of signature, the verifier B computes Verifier B computes

$$K = (y \cdot g^r)^s \mod p \qquad (3)$$

and checks if $h(K, m) = r$ is true.

## 2.2 Zheng's Signcryption Scheme

Y Zheng [12] proposed the following scheme. In this scheme signer parameters are:
  $p, q, g, h(.), g$ are same as previously defined
  KH:      a keyed one way hash function
  (E, D):  the encryption and decryption function of symmetric key cipher.
  $(x_B, y_B)$: B's private and public key

**Signcryption process**

A carries out the following steps to send to a message $m$ to B –
1. Choose $k \in Z_q^*$. let $K = h(y_b^k \mod p)$ and splits K in to $K_1$ and $K_2$ of appropriate length.
2. $r = KH_{K_2}(m, \text{bind\_info})$ where *bind\_info* is an identity of receiver B.
3. $S = \frac{k}{r + x_A} \mod q$
4. $c = E_{K_1}(m)$ and sends to B signcrypted text $(c, r, s)$.

**Unsigncryption**
B computes the following operation to recover $m$ from $(c, r, s)$.
1. Calculates $K = h[(y \cdot g^r)^{s \cdot x_B} \mod p]$ and splits K into $K_1$ and $K_2$
2. Computes $m = D_{K_1}(c)$ and check if $KH_{K_2}(m, bind\_info) = r$ to accept a valid original message.

## 2.3 Blind Signature
Let V denotes A's view of an execution of the signing protocol and let $(m, Sig(m))$ denotes the message signature pair generated in that particular execution. Then a signature scheme is called Blind if A's view and signature pair $(m, Sig(m))$ are statistically independent.

## 3. Our Contribution

### 3.1 Modified Blind Signature using SDSS
The following protocol is a blind version of SDSS.

**Signing on a message m**
1. A randomly chooses $\tilde{k} \in Z_q^*$ and computes $z = g^{\tilde{k}} \mod p$
2. If $gcd(z, q) = 1$. A sends z to B else goes back to 1.
3. B checks if $gcd(z, q) = 1$ and proceed further. Else returns z to A.
4. B randomly chooses $u, \alpha, \beta$ and computes
$$r = h(m, g^u \mod p) \quad (4)$$
$$T = z^r \cdot z^\beta \cdot g^\alpha \quad (5)$$
$$\bar{r} = r + \beta \quad (6)$$
5. B checks if $gcd(\bar{r}, q) = 1$. If yes he sends $\bar{r}$ to A, else goes back to 4 otherwise.
6. A computes $\bar{s} = x_A + \bar{r} \cdot \tilde{k} \mod q$ and forwards it to B.
7. B computes $s = \dfrac{u}{r + \bar{s} + \alpha} \mod q$

**Verification of signature**
Verifier C computes $K = (y_A \cdot T \cdot g^r)^s \mod p$ and checks if $h(m, K) = r$ is true.

***Theorem 1.*** *The pair $(r, s)$ is a valid signature of message m for the SDSS and the above protocol is a blind signature scheme.*

Proof. The validity of the signature pair c may be easily shown as follows
$$(g^r \cdot y_A \cdot T)^s$$
$$= (g^r \cdot g^{x_A} \cdot z^r \cdot z^\beta \cdot g^\alpha)^s \mod p$$
$$= (g^{r+x_A} \cdot z^{r+\beta} \cdot g^\alpha)^s \mod p$$
$$= (g^{r+x_A} \cdot g^{\tilde{k}\cdot(r+\beta)} \cdot g^\alpha)^s \mod p$$
$$= g^u \mod p = K$$
It follows $h(m, K) = r$. Thus $(r, s)$ is a valid signature of m. In order to prove that above protocol is blind we have to show that any given view V and a valid message signature tuple $(m, Sig(m))$ with $r \equiv 0 \mod q$, there exist a unique pair of blinding factors $\alpha$ and $\beta$. Because $\alpha$ and $\beta$ are chosen randomly, the blindness of the scheme follows –

If $(r, s)$ is a signature tuple for the message m, generated during execution of the protocol with view V consisting of $\tilde{k}, z = g^{\tilde{k}} \mod p, \bar{r}$ and $\bar{s} = x_A + \bar{r} \cdot \tilde{k} \mod q$. Then the following equation must hold for $\alpha$ and $\beta$.
$$T = z^r \cdot z^\beta \cdot g^\alpha$$
$$\bar{r} = r + \beta$$

$$s = \frac{u}{r + \bar{s} + \alpha} \bmod q$$

because $z$, $r$, $\bar{r}$, are relatively prime to $q$, the blinding factors $\alpha$ and $\beta$ are uniquely determined by the first two equations:

$$\alpha = s^{-1}.u - (r + \bar{s}) \bmod q$$
$$\beta = \bar{r} - r$$

By substituting $\bar{s} = x_A + \bar{r} \cdot \bar{k} \bmod q$, we obtain

$$\begin{aligned}
z^r \cdot z^\beta \cdot g^\alpha &= g^{(\bar{k}.r + \bar{k}.\beta) + \alpha} \\
&= g^{\bar{k}.r + \bar{k}.(\bar{r} - r) + s^{-1}.u - (r + \bar{s})} \\
&= g^{\bar{k}.\bar{r} + s^{-1}.u - r - \bar{s}} \\
&= g^{\bar{k}.\bar{r} - \bar{s}} g^{s^{-1}.u} g^r \\
&= g^{-x_A} g^{s^{-1}.u} g^r \\
&= g^{s^{-1}.u - x_A + r} \\
&= g^{s^{-1}.u} \cdot g^{-x_A} g^r \\
&= g^{s^{-1}.u} y_A^{-1} g^r = T \bmod p
\end{aligned}$$

it gives $(y_A.T.g^r)^s = g^{s^{-1}.u.s} = g^u = K$. which is true.

An interesting characteristic of above stated shortened blind signature scheme is that $g^u \bmod p$ is not explicitly contained in a signature $(r, s)$. It can be recovered from the pair $(r, s)$ and other public parameters. This motivates us to construct a blind signcryption scheme in way similar to Zheng's signcryption.

### 3.2 Blind Signcryption

In this section we propose our blind Signcryption scheme. In describing our scheme, we will use E and D to denote the encryption algorithm and decryption algorithm. Encrypting a message with a key K is indicated as $E_K(m)$ while decrypting a ciphertext is denoted by $D_K(m)$. In addition we use $KH_K(m)$ to denote hashing a message $m$ with a keyed hash algorithm KH under a key K.

Assume that A has chosen a secret key $x_A$ from [1, 2, ..., q] and made public her matching public key $y_A = g^{x_A} \bmod p$. Similarly B and C's secret keys are $x_B$ and $x_C$, and their respective public keys are $y_B = g^{x_B} \bmod p$ and $y_C = g^{x_C} \bmod p$.

The signcryption and unsigncryption protocol using a modified blind signature are remarkably simple. To blindly signcrypt a message m we adopt the following protocol:

**Signcryption Process**
1. A randomly chooses $\tilde{k} \in Z_q^*$ and computes $z = g^{\tilde{k}} \bmod p$
2. A sends $z$ to B if $\gcd(z, q) = 1$ else he goes back to step1
3. B randomly chooses $\alpha$, $\beta$ and computes $T = z^r z^\beta g^\alpha$
4. Chooses $u \in Z_q^*$. Let $K = h(y_C^k \bmod p)$ and splits K in to $K_1$ and $K_2$ of appropriate length.
5. B computes $\bar{r} = r + \beta$, $c = E_{K_1}(m)$ and $r = KH_{K_2}(m, bind\_info)$ where $bind\_info$ is an identity of receiver C, and sends $\bar{r}$ to A.
6. A signs $\bar{r} = x + \bar{r}\,\bar{k}$ and sends $\bar{s}$ to B.
7. B computes $s = \dfrac{u}{r + \bar{s} + \alpha}$ (mod $q$) and sends to C the signcrypted text $(c, r, s)$.

**Unsigncryption**
C computes the following operation to recover the original message $m$ from $(c, r, s)$.
1. Calculate $K = h[(y.T.\,g^r)^{s\,x_C} \bmod p\,]$
2. Split K into $K_1$ and $K_2$

3. $m = D_{K1}(c)$
4. Check if $KH_{K2}(m, bind\_info) = r$ to accept *m* as a valid original message.

## 4. Security Analysis

The blind signcryption scheme satisfies all the properties for an authenticated encryption. It provides anonymity, untracebility of participants, unlinkablity, confidentiality, authentication and non-repudiation. anonymity, untracebility of participants and unlinkablity are the properties which are due blindness of the protocol. This may be similar to theorem 1 discussed in section 3.1 Confidentiality is achieved by using encryption algorithm EK. The authentication is achieved by having the signature for which sender uses it's private key to signcrypt. In case of dispute, the zero knowledge proof protocol between the third party judge and recipient, may settle their problem. That is, Signer can not deny that he made that signcrypted text with the recipient. Most of the security discussion is similar to the original schemes [13, 3, 8, 14].

## 5. Conclusion

Blind signature schemes enable a useful protocol that guarantee the anonymity of the participants while Signcryption offers authentication of message and confidentiality of messages at the same time and more efficiently. In this paper, we present a blind signcryption scheme that combines the functionality of blind signature and signcryption. This blind Signcryption is useful for applications that are based on anonymity untracebility and unlinkability.